\documentclass[preprint2,twoside]{hwo}

\usepackage{graphicx}

\input{hwo.h}

\begin{document}

\title{\textbf{\LARGE The role of the \emph{Hubble Space Telescope} in advancing our understanding of atmospheric escape in exoplanets}}

\author{
Leonardo~A.~Dos~Santos,$^{1,2,\dagger}$ 
Arika~Egan,$^{3}$
Kevin~France,$^{4}$
Eric~Gaidos,$^{5}$
Antonio~Garc\'ia~Mu\~noz,$^{6}$
R.~O.~Parke~Loyd,$^{7}$
Keighley~Rockcliffe,$^{8}$
Mercedes~López-Morales,$^{1}$
P.~Christian~Schneider,$^{9}$
Arif~Solmaz,$^{10}$
Michael~Zhang,$^{11}$
Vladimir~S.~Airapetian,$^{12}$
Munazza~K.~Alam,$^{1}$
Laura~N.~R.~do~Amaral,$^{13}$
Tommi~Koskinen,$^{14}$ %
Seth~Redfield,$^{15}$
Jake~D.~Turner$^{16}$
}
\affil{$^1$\small\it Space Telescope Science Institute; $^\dagger${\tt ldsantos@stsci.edu}; $^2$Department of Physics and Astronomy, Johns Hopkins University; $^3$Johns Hopkins Applied Physics Laboratory; $^4$Laboratory for Atmospheric and Space Physics, University of Colorado Boulder; $^{5}$University of Hawai'i at M\={anoa}; $^6$Universit\'e Paris-Saclay, Universit\'e Paris Cit\'e, CEA, CNRS, AIM; $^7$Eureka Scientific, Inc.; $^8$NASA Goddard Space Flight Center/CRESST; $^{9}$Kiel University; $^{10}$Department of Mechatronics Engineering, Istanbul Health and Technology University; $^{11}$University of Chicago; $^{12}$NASA Goddard Space Flight Center/SEEC; $^{13}$Arizona State University; $^{14}$University of Arizona; $^{15}$Wesleyan University; $^{16}$Department of Astronomy and Carl Sagan Institute, Cornell University}

\author{\small{\bf Endorsers of this white paper are listed at the end of the document.} }

\begin{abstract}
An important evolutionary pathway for planetary atmospheres is escape to space, which has been studied on Earth and Mars for several decades and more recently in exoplanets. A particularly important regime is the hydrodynamic escape, wherein atmospheric mass escapes the planet at high rates in a collisional fluid outflow. This process is used to partly explain the early evolution of rocky planets in and out of the Solar System, as well as key aspects of exoplanet demographics. Hydrodynamic escape is not occurring in the Solar System planets, so our only option for such observations is through exoplanets. The ultraviolet (UV) capabilities of the \emph{Hubble Space Telescope} (\emph{HST}) are fundamental to detect hydrodynamic escape and measure the resulting mass-loss rates for a range of planetary systems and to identify targets for surveys with the Habitable Worlds Observatory. We discuss here what kinds of observations and instrument modes are necessary to continue studying atmospheric escape in exoplanets for the next decade, as well as how to advance our understanding of planetary evolution and habitability. \textit{This article is a response to the call for white papers by the Space Telescope Science Institute on ``Building a Roadmap for Hubble science into the 2030s."}
  \\
  \\
\end{abstract}

\vspace{2cm}

\section{Introduction}

Hydrodynamic escape is the dominant process that erodes atmospheres of planetary bodies that lack abundant molecular coolants and have appreciable high-energy irradiation from the host star \citep{Gross1972}. This process was initially invoked to explain the early evolution of the atmospheres of Venus, Earth \citep{Watson1981, Yung1989} and Mars \citep{Hunten1987}. Currently, no planetary body in the Solar System experiences hydrodynamic escape \citep{Gronoff2020}. The discovery of transiting exoplanets \citep{Charbonneau2000, Henry2000} enabled us to observe the atmospheres of worlds that are currently experiencing hydrodynamic escape. The most direct technique utilized to observe this process is transmission spectroscopy, in which the observer measures time- and wavelength-dependent changes in the spectrum of a star when a planet and its atmosphere transit the stellar disk \citep{Seager2000, Charbonneau2002}.

Although there have been important contributions to the study of exoplanets from ground-based telescopes \citep[e.g.,][]{Vogt1994, Cosentino2012, Quirrenbach2012, Jensen2012, Allart2018, Espinoza2019, Pepe2021, Bouchy2025, Snellen2025}, space-based observatories, such as \emph{HST} and \emph{JWST}, offer unparalleled flux stability over time, sensitivity, and access to photons that would otherwise be absorbed by the Earth's atmosphere -- specifically the vacuum UV, where the relevant spectra features for atmospheric escape are Lyman-$\alpha$ and lines of C, N, O, Si, Mg and Fe \citep[e.g.,][]{Vidal2003, Vidal2004, Sing2019}. These capabilities are paramount to answer key science questions related to the evolution of planetary systems and inform future missions targeting potentially habitable exoplanets.

\section{Key science questions}

\paragraph{How does atmospheric escape affect the demographics and evolution of exoplanets?}

\begin{figure}[h]
    \centering
    \includegraphics[width=0.95\linewidth]{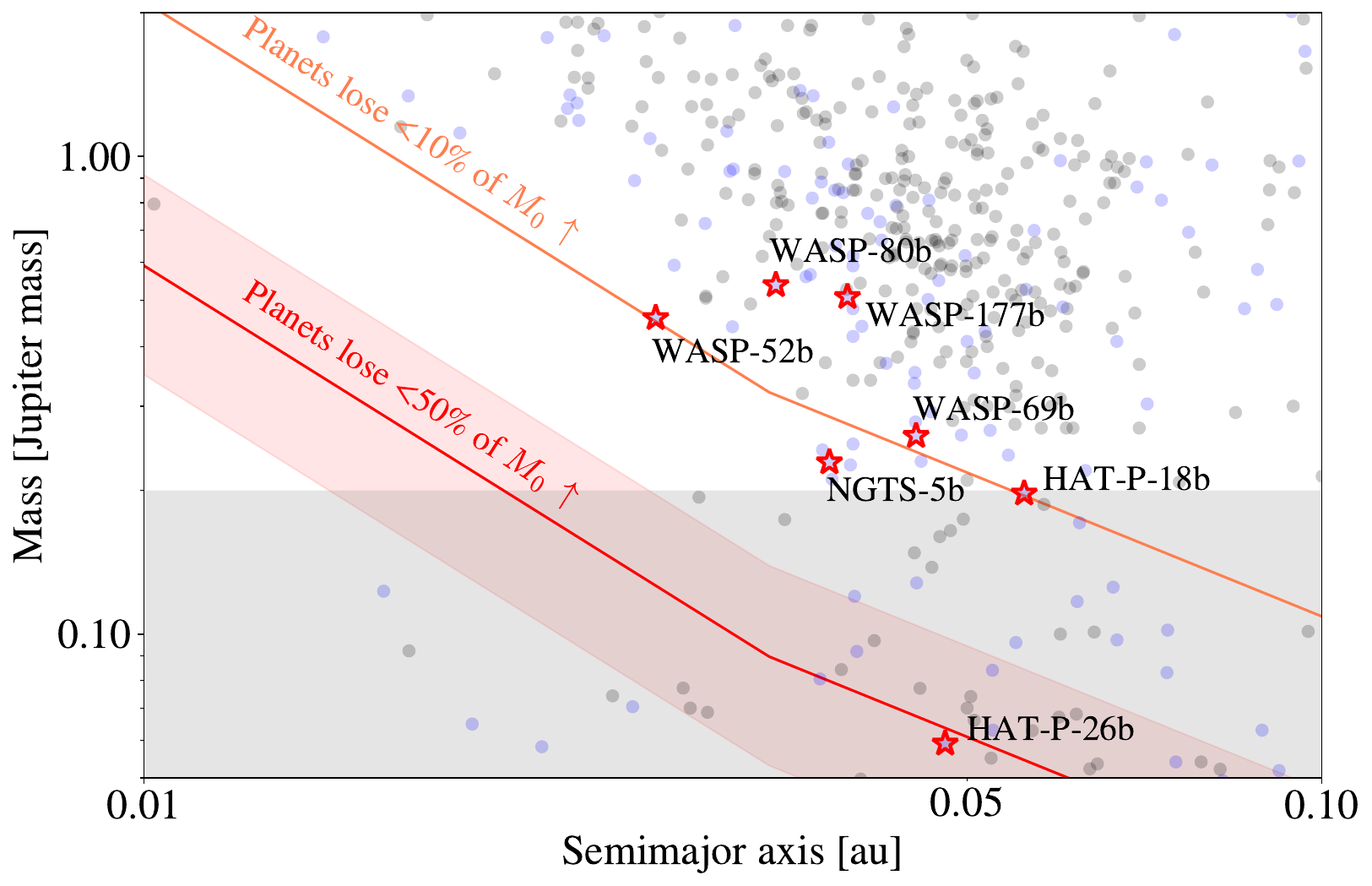}
    \caption{Semi-major axis versus planetary mass diagram for a population of transiting exoplanets. Stars are symbols for the planets observed in a ground-based atmospheric escape campaign. Planets above the red (orange) curve cannot have lost more than 50\% (10\%) of their initial mass $M_0$ to photoevaporation. For more details, see \citet{Vissapragada2022b}.}
    \label{fig:desert}
\end{figure}

Two distinct features in the exoplanet population may be largely shaped by hydrodynamic escape. The first is known as the hot-Neptune desert \citep[e.g.,][]{Howard2010, Szabo2011, Beauge2013, Mazeh2016}, originally thought to be carved by atmospheric escape \citep[e.g.,][]{Jackson2012, Lopez2013, Kurokawa2014}, Roche-lobe overflow \citep[e.g.,][]{Vissapragada2025, Hallatt2026}, high-eccentricity tidal migration \citep{CGonzales2026} or planet formation \citep{Batygin2016, Boley2016}. A recent observational study by \citet{Vissapragada2022b} has shown that the upper edge of the hot-Neptune desert is stable against atmospheric loss \citep[except for, perhaps, ultrahot Jupiters;][]{Garcia2019}, and that a combination of the effects above is needed to explain it \citep[][see Figure \ref{fig:desert}]{Owen2018}. And yet, observational surveys have slowly been populating the desert \citep[e.g.,][]{CGonzales2024, Cui2026}, so the emergence of this feature may depend on other aspects than just planet radius and semi-major axis, i.e. planet composition, and perhaps the early history of stellar activity. Answering this question will require us to measure the mass-loss rates of a large sample of exoplanets.

\begin{figure*}[t]
    \centering
    \includegraphics[width=0.95\linewidth]{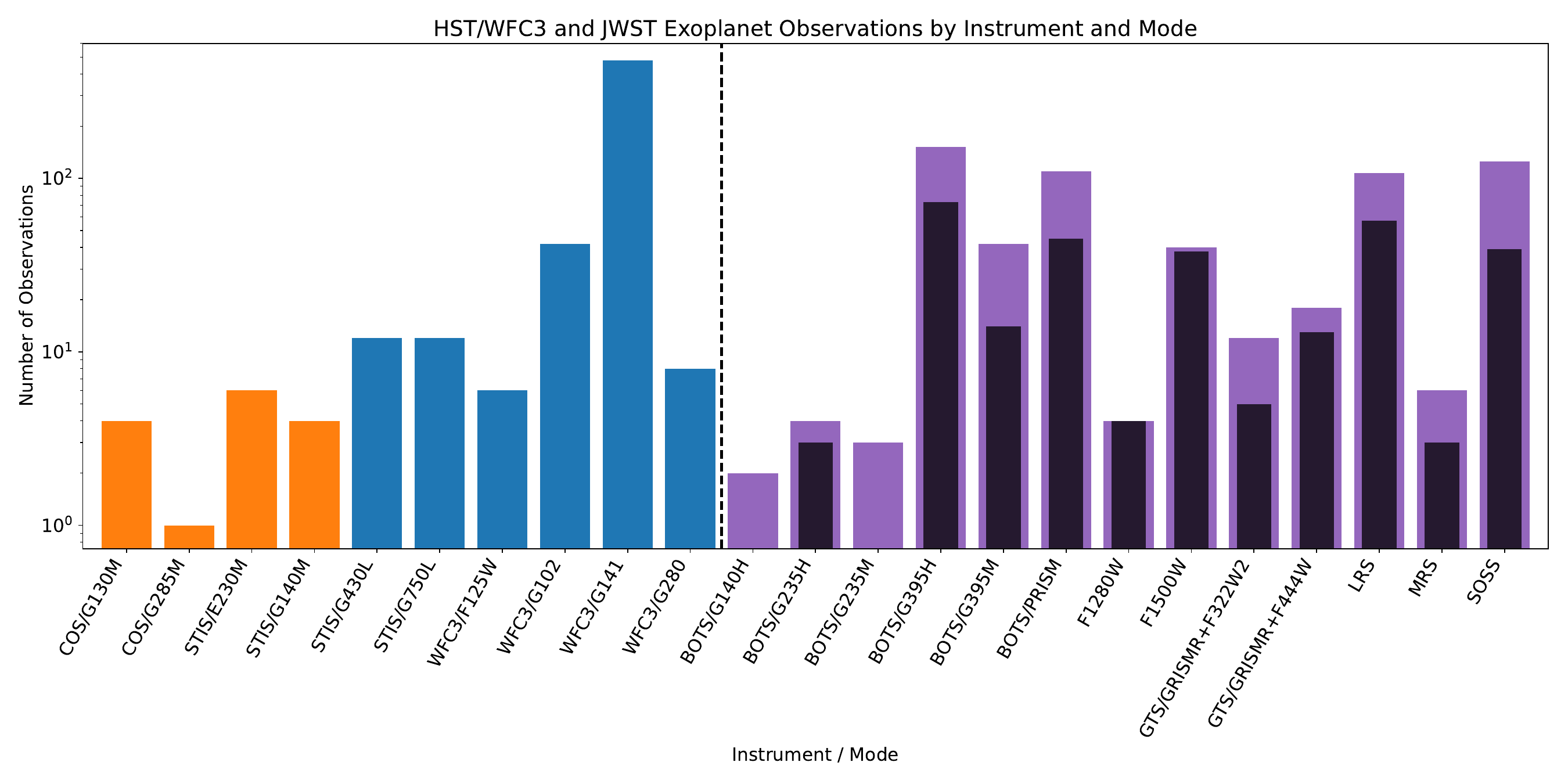}
    \caption{Histogram of archived exoplanet transit observations obtained with \emph{HST} (left of the dashed divider) and \emph{JWST} (right), grouped by instrument and observing mode. \emph{HST} UV configurations are highlighted in orange, while optical and infrared \emph{HST} modes are shown in blue and \emph{JWST} modes in purple. The black overlaid histogram denotes the subset of \emph{JWST} targets with bulk densities $\rho > 2.97\,\mathrm{g\,cm^{-3}}$. Data come from TrExoLiSTS \citep{Nikolov2022} and MAST.}
    \label{fig:modes}
\end{figure*}

Simulations and observational campaigns have revealed a bimodal distribution of planets near 1.8~R$_\oplus$ \citep{Owen2013, Owen2017, Fulton2017, Fulton2018}. This second demographic feature, known as the radius gap, is thought to divide the population of sub-Neptunes into those that still retain H/He envelopes (mini-Neptunes) to those that do not (super-Earths)\footnote{\footnotesize{These are informal definitions, as different authors use some of these terms interchangeably.}}. This feature is hypothesized to be carved by photoevaporation \citep{VanEylen2017}, core-powered mass loss \citep{Ginzburg2018, Gupta2019, Owen2023b}, or a direct product of planet formation \citep{Zeng2019, Venturini2020, Lee2022}. Recently, this picture became even more complicated by the hypothesis that some mini-Neptunes could be water worlds \citep[e.g.,][]{Madhusudhan2020, Luque2022, Piaulet2024}. Direct evidence for transformation of sub-Neptunes into super-Earths with time has been obtained for a restricted demographic \citep[M dwarf planets;][]{Gaidos2024}, and direct observations of escape with space telescopes among small exoplanets are scarce (see Figure \ref{fig:modes}). So, the overall role of atmospheric escape in the evolution of super-Earths, mini-Neptunes and water worlds remains poorly constrained \citep[e.g.,][]{Garcia2020, Garcia2021, Gupta2025}.

\paragraph{What high-energy irradiation do planets receive, and how efficiently is it converted into an outflow?}

The ratio of the gain in gravitational potential energy of escaping gas to the ionizing energy deposited into the atmosphere by the host star is known as the photoevaporation efficiency \citep{Erkaev2007, Salz2016, 2026Frelikh}. Simulations predict that planets with deeper gravitational potentials tend to have lower efficiencies \citep{Koskinen2014, Caldiroli2022}, but it may also depend on other factors such as composition \citep{Yoshida2025}.  In reality we have not yet accurately measured this key parameter for a sample of planets \citep[e.g.,][]{McCreery2025}. In addition, knowledge about the stellar extreme-UV (EUV) spectrum is required, but we lack observational constraints for stars beyond a few parsecs (see Section \ref{sec:hwo} for more details). The photoevaporation efficiency and EUV flux will modulate mass-loss rates over time and thus have a significant impact on the evolution of planets at the demographic scale. Measuring mass-loss rates for a large sample and comparing them with theoretical predictions will enable us to estimate photoevaporation efficiencies; these predictions will be particularly important for small exoplanets, which might have very diverse atmospheric compositions.

\paragraph{What is the mass-loss rate of young exoplanets?}

Models predict that sub-Jovian exoplanets will experience higher mass loss rates during the first $\sim$100 Myr -- 1 Gyr because they are inflated from the initial entropy of formation, and because the host star is emitting more high-energy radiation \citep[e.g.,][]{Jackson2012, Owen2013}. Although the vast majority of known transiting planets orbit older stars, the advent of the \emph{Gaia} mission \citep{Gaia2016}, which has allowed us to identify many more nearby young stars, and the all-sky \emph{TESS} transit survey that has monitored most of them \citep{Ricker2015}, has led to a steadily increasing number of young transiting planets and observations to search for escaping atmospheres  \citep[e.g.,][]{Hirano2020,Gaidos2020,Zhang2022, Gaidos2022, Gaidos2023, Zhang2023, Rockcliffe2023}. While challenges remain, such as the elevated activity of young stars changing their spectrum during transits, as well as the greater distances to young stellar clusters, continued monitoring by \emph{TESS} and planned/possible new missions such as PLATO \citep{Rauer2025} and EVE \citep{Howard2025} will continue to expand the sample of accessible young transiting planets. 

\paragraph{How does chemical composition affect the evolution of planets and vice versa?}

Studies on Solar System planets infer that atmospheric escape plays a major role in changing the chemical composition of planetary-mass bodies \citep{Ringwood1966, Watson1981, Hunten1987, Yung1989}. In the case of giant planets, it seems that photoevaporation has a limited role in their evolution  \citep{Louca2023} and their chemical composition is likely set by formation pathways \citep[e.g.,][]{Oberg2011, Mordasini2016}. On the other hand, sub-Jovian worlds -- the most common type of planet in the solar neighborhood -- are predicted to be more susceptible to atmospheric chemical composition changes \citep{Louca2025}, becoming more metal rich after losing mostly H and He in their early lives. There is a negative feedback that has not been thoroughly studied, since the increase in metallicity also tends to decrease mass-loss rates due to the increase in mean molecular weight and increased thermospheric cooling \citep[e.g.,][]{garciamunozetal2024,Vissapragada2024, Zhang2025, Yoshida2025,Taylor2026}. Thus, it is possible that sub-Jovian planets might be more resilient to atmospheric escape than predicted by pure H+He models. Testing this hypothesis will require more observations of escaping H, He and metals for these kinds of planets as well as knowledge of the compositions of their upper atmospheres.

\paragraph{What is the impact of atmospheric escape on habitability?}

Planets must retain an atmosphere in order to be habitable, which is not a foregone conclusion \citep[e.g.,][]{Segura2010, OMJames2017, Estrela2020}. \citet{Zahnle2017} proposed the ``cosmic shoreline" as a framework to predict which planets could potentially retain atmospheres \citep[see also][]{2025Ji}; observational campaigns with \emph{JWST} have already started exploring this hypothesis \citep[e.g.,][]{Taylor2025, Glidden2025, Espinoza2025, Allen2026}. Modeling of individual planets, particularly with M dwarf hosts, have suggested that some of them are less likely to retain atmospheres \citep[e.g.,][]{2020France, 2025Amaral, 2026Brain}. UV observations of exoplanet hosts over time with \emph{HST} are crucial to determine the habitability of a given system, as they are an input to atmospheric escape and evolution models. The ongoing \emph{JWST}/\emph{HST} Rocky Worlds DDT Program, in addition to studying atmospheres of rocky planets, will provide valuable information for models of atmospheric loss \citep{Redfield2024}.

\section{Relevant \emph{HST} observing modes and capabilities}

The main \emph{HST} instruments for atmospheric escape observations are the Space Telescope Imaging Spectrograph \citep[STIS;][]{Vidal2003, Lecavelier2010, Ehrenreich2015, Sing2019, DSantos2020b, Zhang2022, Gressier2023, Loyd2025, Baldwin2026} and the Cosmic Origins Spectrograph \citep[COS;][]{Fossati2010, Linsky2010, Bourrier2018b, DSantos2019b, Garcia2021, BJaffel2022, DSantos2023}. Together, they cover wavelengths from as low as 912~\AA\ (albeit with low sensitivity at $\lambda < 1100$~\AA) to $3000$~\AA\ at medium- to high-resolution (M and H) modes. Spectral resolving powers higher than $R \sim 10\,000$ are usually necessary to resolve the in-transit absorption, which helps in discerning outflow velocities and distinguishing planetary from stellar signals. High-resolution observations enable better characterization of the geocoronal contribution, local ISM absorption, and intrinsic stellar emission \citep{Wood2005}. In some cases, such as spectral features that are isolated or the magnetically quiet host stars, lower-resolution (L) modes are acceptable, especially if high signal-to-noise ratios are required or if there is a forest of weak spectral features. L modes on \emph{HST} can cover wavelengths as long as $\lambda < 10\,000$~\AA\ with STIS or the Wide-Field Camera 3 \citep[WFC3; e.g.,][]{Gascon2025, Boehm2025}.

In the FUV, STIS is used most often for observations of neutrals -- namely Lyman-$\alpha$ (H\,{\sc i}) and oxygen (O\,{\sc i}) -- since the slit design allows for a reliable subtraction of the geocoronal contamination. However, in the cases where the Lyman-$\alpha$ or O\,{\sc i} emission lines are bright enough, it is possible to remove the geocoronal contamination through template fitting \citep{Bourrier2018, DSantos2019b, CAguirre2023}. The COS instrument is better suited to the observation of ionized lines and fainter features due to its increased sensitivity, spectral resolution and wider wavelength coverage than the medium-resolution modes of STIS. If the source is too bright for COS, then STIS is the fallback instrument due to its higher resilience to higher count rates in the FUV. In the NUV, STIS is the most used instrument due to its wavelength coverage. Where high-resolution spectroscopy is not required, the STIS grating G230L is a viable observing mode at these wavelengths. More recently, the WFC3/UVIS mode has been used to obtain low-resolution NUV-optical transmission spectra of transiting planets as well, which can trace some escaping species \citep{Wakeford2020, Lothringer2022}. A summary of relevant \emph{HST} modes for observations of atmospheric escape is in Table \ref{tab:modes}.

An impediment to some FUV observations has been the policy intended to protect against variable M dwarfs \citep{osten_bright_2017}, which pose a risk of exceeding the brightness limits of \emph{HST}'s UV detectors due to unpredictable flares. This barrier could be substantially lowered, while still preserving detector safety, by reevaluating the flare protection policy to utilize the now large database of M dwarf flare observations at UV wavelengths. This change to operation constraints could enable new UV transit observations of planets orbiting bright and/or active M dwarfs.

\begin{table*}[h]
    \centering
    \caption{Relevant instrument modes for exoplanet atmospheric escape observations.} \label{tab:modes}
    \begin{tabular}{l|ccc}
    \hline
    Spectral features & Instrument & Detector & Gratings \\
    \hline
    H\,{\sc i}, O\,{\sc i} (most science cases) & STIS & FUV-MAMA & G140M, E140M, E140H \\
    H\,{\sc i}, O\,{\sc i} (specific cases) & COS & FUV-XDL & G130M \\
    FUV continuum (most science cases) & STIS & FUV-MAMA & G140L \\
    FUV continuum (specific cases) & COS & FUV-XDL & G140L \\
    Mg\,{\sc i}, Mg\,{\sc ii}, Fe\,{\sc i}, Fe\,{\sc ii} & STIS & NUV-MAMA & G230M, E230M, E230H \\
    Other metal lines (e.g., Si, C, N, S) & COS & FUV-XDL & G130M, G160M, G140L \\
    NUV continuum & STIS & NUV-MAMA & G230L \\
    NUV-optical continuum & WFC3 & UVIS-CCD & G280 \\
    \hline
    \end{tabular}
\end{table*}

\section{Supporting the mission of the Habitable Worlds Observatory} \label{sec:hwo}

The Habitable Worlds Observatory (HWO) is poised to detect Earth-like exoplanets by direct imaging and spectroscopy \citep[e.g.,][]{NAP26141, Arney2026, Berdyugina2026}, but will also continue to explore transiting systems \citep[e.g.,][]{DSantos2025, Cubillos2026, Wakeford2026} and a wide range of cooler worlds by direct techniques \citep[e.g.,][]{Min2026, Hu2026}. A full characterization of these planets will involve models of their evolution through atmospheric escape, particularly for mini-Neptunes and rocky planets.

Photoevaporation is driven mostly by energetic photons with wavelengths shorter than 911~\AA\ \citep{Lammer2003}. Since most stars do not have measurements of their EUV spectra due to a lack of a sensitive EUV astrophysics observatory and ISM absorption, often our best option is to reconstruct their EUV spectra using a combination of X-ray and UV measurements \citep[e.g.,][]{Duvvuri2021}. Thus, characterizing potential HWO target stars with \emph{HST} today will enable a deeper understanding of the EUV irradiation of any planets. The definition of the HWO ``Living Worlds'' target sample \citep{Tuchow2025} highlights the importance of understanding the high-energy radiation (including UV) from nearby stars in order to model atmospheric evolution and habitability of rocky exoplanets. However, only 33 of the 164 stars in one HWO target sample possess reliable space-based UV measurements \citep{Harada2024}.

These limitations directly affect atmospheric escape and habitability studies. UV emission lines such as Ly$\alpha$, Mg\,{\sc ii}, C\,{\sc ii}, Si\,{\sc iii}, Si\,{\sc iv}, N\,{\sc v}, and O\,{\sc i} trace the chromospheres and transition regions of host stars, while Fe\,{\sc xii}, Fe\,{\sc xix}, and Fe\,{\sc xxi} trace the corona and are routinely used to reconstruct the currently unobservable EUV spectrum that powers hydrodynamic escape. Without these measurements, atmospheric escape efficiencies and photochemical models will remain poorly constrained. \citet{Peacock2025} further showed that most archival UV observations consist of isolated visits that do not capture stellar variability, flares, or activity cycles, despite these processes strongly modulating atmospheric mass loss, ionization structure, and abiotic chemistry in planetary atmospheres.

\emph{HST} remains the only operating observatory capable of obtaining moderate- and high-resolution UV spectroscopy of nearby exoplanet host stars before HWO begins operations. Coordinated surveys with \emph{HST} could establish panchromatic stellar spectral energy distributions, measure flare statistics and long-term variability, and calibrate stellar atmosphere models used in atmospheric escape and photochemical simulations \citep{Harada2024, Peacock2025}. To maximize the long-term value of such observations, community-led efforts should strive to release standardized stellar-context metadata -- e.g., activity index -- along with the resulting \emph{HST} products.

\section{Future outlook}

\emph{HST} recently began its first-ever survey-scale program focused on atmospheric escape: the Survey of Transiting Exoplanets in Lyman-$\alpha$ \citep[STEL$\alpha$,][]{Loyd2024}. This program will map hydrogen loss across the exoplanet population through observations of Lyman-$\alpha$ transits, enabling the first-ever population-level understanding of atmospheric escape. STEL$\alpha$ provides a solid base of experience for conducting new and complementary large-scale surveys of atmospheric escape with \emph{HST} in the UV domain that is inaccessible from the ground. By following STEL$\alpha$ with one or more UV surveys that broaden coverage in time, wavelength, and/or targets, \emph{HST} can deepen our observational understanding of atmospheric escape. 

In the time domain, \emph{HST} could observe transits over timescales of years to identify changes in atmospheric escape in response to the planet's space environment (XUV irradiation and stellar wind). Solar and stellar XUV emission varies by an order of magnitude over magnetic activity cycles \citep{johnstone_active_2021}. An \emph{HST} campaign could monitor the XUV variations (via FUV proxy emission) of multiple planet-hosting stars with observable outflows, re-observing a transit once changes in XUV emission exceed a set threshold. A coordination with ground-based monitoring of activity tracers in the optical \citep[e.g.,][]{Cincunegui2007} would also be appropriate.

A second large-scale time-domain survey could sample stellar variability on transit timescales, enabling more accurate interpretations of transit signals. Stellar Lyman-$\alpha$ variability is largely unconstrained on the roughly 8~h timescale of typical transit observations with \emph{HST}. \emph{HST} could initiate a survey to constrain stellar variability in Lyman-$\alpha$ and UV metal lines of stars with detected transits to mitigate these uncertainties. A future survey could apply the STEL$\alpha$ survey approach to upper atmospheric metals, which are direct tracers of hydrodynamic escape. A similar campaign is being carried out by current programs \citep{DSantos2025propa, DSantos2025propb}, although at a smaller scale. A large program could systematically survey systems predicted to have observable metal escape, or where non-detections would be particularly constraining, mapping mass-loss rates throughout a wide swath of the exoplanet population, particularly for those for which Lyman-$\alpha$ observations are unfeasible.

There is a growing body of evidence that planetary outflows can produce signals many hours outside of transit, meaning a number of existing observations sampling only near the planet’s optical transit might never have measured an accurate baseline \citep[e.g.,][]{Allart2025}. Thus, we encourage investigators to explore observing strategies that ensure the obtaining of accurate flux baselines well out-of-transit. Finally, many of planets with detections of atmospheric escape do not have a full characterization of their host star EUV spectra, so modelling efforts have to rely on proxy spectral energy distributions \citep[e.g.,][]{DSantos2022pwinds}, leading to less accurate interpretations. Such efforts would significantly benefit from directly measuring the far- to near-UV spectra of stars that host photoevaporating exoplanets while \emph{HST} is still operational.

\paragraph{Acknowledgments.} We thank the participants of the STScI Fall Science Workshop 2025, ``Atmospheric Escape and Replenishment in Planetary Systems," for the many discussions that inspired the writing of this white paper.

\newpage

\paragraph{Endorsed by:} 
Adina D.\, Feinstein (Michigan State University), 
Akash Gupta (Princeton University),
Alain Lecavelier des Etangs (Institut d'Astrophysique de Paris),
Alfred Vidal-Madjar (Institut d'Astrophysique de Paris),
A.~A.~Vidotto (Leiden University),
Allison Youngblood (NASA GSFC),
Antonija Oklop\v{c}i\'{c} (University of Amsterdam), 
Ava Morrissey (Carnegie Science Observatories),
Chenliang Huang (Shanghai Astronomical Observatory),
Elijah Mullens (Cornell University), 
Ethan Schreyer (University of California Santa Cruz),
Hannah Diamond-Lowe (STScI),
Hilke E.\, Schlichting (University of California Los Angeles), 
Joshua D.\,Lothringer (STScI),
Julia V.\, Seidel (Observatoire de la Côte d'Azur), 
Katherine A. Bennett (Johns Hopkins University),
Lakeisha M. Ramos Rosado (Johns Hopkins University),
Luca Fossati (Space Research Institute, Austrian Academy of Sciences), 
Matthew M.\,Murphy (Michigan State University),
Morgan Saidel (California Institute of Technology), 
Natalie H. Allen (Johns Hopkins University),
Patricio E. Cubillos (Space Research Institute, Austrian Academy of Sciences),
Romain Allart (Universit\'e de Montr\'eal),
Shreyas Vissapragada (Carnegie Science Observatories),
Tyler Richey-Yowell (McDonald Observatory, The University of Texas at Austin),
Vincent Bourrier (Université de Genève),
Yao Tang (University of California Santa Cruz).

\bibliography{author.bib}

\end{document}